\documentclass[useAMS,usenatbib,usegraphicx]{mn2e}

\title[Kinematics of a compact ETG at $z\simeq1.4$]{LBT-LUCIFER spectroscopy: kinematics of a compact early type 
 galaxy at $z\simeq1.4$
\thanks{Based on observations made at the Large Binocular Telescope (LBT) at Mt. Graham (Arizona, USA).
}}
\author[M. Longhetti, P. Saracco, A. Gargiulo, S. Tamburri, I. Lonoce]{M. Longhetti$^{1}$\thanks{E-mail:
marcella.longhetti@brera.inaf.it (OAB)}, P. Saracco $^{1}$, A. Gargiulo$^{1}$, S. Tamburri$^{1}$, I. Lonoce$^{1}$
\footnotemark[1]
\\
$^{1}$ INAF-Osservatorio Astronomico di Brera, via Brera 28, 20121 Milano, Italy}
\begin{document}

\date{Accepted 2010 December 15. Received 2010 October 14; in original form 2010 October 11}

\pagerange{\pageref{firstpage}--\pageref{lastpage}} \pubyear{2010}

\maketitle

\label{firstpage}

\begin{abstract}
We present a high signal to noise (S/N$>$10)  medium resolution (R=2000) LBT-LUCIFER spectrum of the early-type galaxy
(ETG) S2F1-142 at $z\simeq1.4$. 
By means of the CaT line at $8662$ \AA, we measured its redshift $z=1.386\pm 0.001$ and we estimated
its velocity dispersion
$\sigma_{v}=340  ^{-60}_{+120}$ km/s. Its corresponding virial mass is 3.9$\times10^{11}$ M$_\odot$,
compatible with the stellar mass estimates obtained assuming Initial Mass Functions (IMFs) 
 less dwarf rich than the Salpeter one.
S2F1-142 is a compact galaxy with $R_{e}$=3.1$\pm$0.2 kpc, i.e., an effective radius 
more than three times smaller than the average $R_{e}$ of early-type galaxies with the same mass in the local universe.
At the same time, we found local and high redshift galaxies with a similar mass content and  similar effective radius
confirming that it is fully consistent with the already available measures of $R_{e}$ and $\sigma_{v}$ both in
the local and in the distant universe. 
Considering the distribution of $R_{e}$ and $\sigma_{v}$ as a function of the stellar mass content of ETGs, both in the local
and in the distant universe,  we noticed that
the measured velocity dispersions of the more compact galaxies are on average slightly lower 
than expected on the basis of their compactness and the virial theorem,
suggesting that
{\it i)} their dark matter content is lower than in the more diffuse galaxies 
and/or {\it ii)} their luminosity profiles are steeper than in the more diffuse galaxies and/or
  {\it iii)} their larger compactness is an apparent effect caused by the overestimate of their
stellar mass content (due to bottom lighter IMF and/or systematic affecting the stellar mass estimates). 
%The latter point could be the case if more compact galaxies would have 
%a bottom lighter IMF than more diffuse galaxies or if some of the many systematic affecting the stellar mass estimates 
%is artificially growing their stellar mass.

\end{abstract}

\section{Introduction}
Understanding the formation of early-type galaxies (ETGs) is one of the crucial issues of 
cosmology, since they contain most of the present-day stars and baryons (e.g.
\citealt{ren}, \citealt{fuk}). After their first spectroscopic detection at $z>1.5-2$ 
(\citealt{dun}, \citealt{spin}, \citealt{s3}, \citealt{mc}, \citealt{cim04}, \citealt{gla}),
only in the last 10 years
many estimates of their physical properties have been possible, e.g., age and metallicity of their stellar content
(\citealt{gobat}, \citealt{l5}, \citealt{daddi}), 
morphological 
profiles  and color gradients (\citealt{gar12}, \citealt{gar11}, \citealt{mcg}, \citealt{moth}).
The emerging picture seems to tell us that most of
them formed the bulk of their stars at $z>2-3$, even if further small star forming events happened during their successive
evolution (e.g., \citealt{ren}). Furthermore,  observations of $z\simeq 1.5$ massive ETGs show evidence 
of both compact (i.e., $R_{e}$ a factor 3-6 smaller than local massive galaxies) and more diffuse
galaxies (\citealt{s10}, \citealt{man10}),
and the evolution  of the most compact ETGs from $z>1.5$ to $z=0$ in the sense of an enlarging
of their effective radii is still a debated possibility (\citealt{s9}; \citealt{ono}; \citealt{man10}).
Within this picture, a still missing measure of high redshift ETGs properties is their velocity dispersion $\sigma_{v}$,
up to now available only for few $z>1.2$ galaxies and even for fewer compact galaxies (i.e., \citealt{vdk}; \citealt{toft}; \citealt{vds}).
The measure of the velocity dispersion allows to tackle the issue of ETGs formation and evolution from the point of view
of their dynamics, and to estimate their total mass content, a primary parameter in galaxy formation models.

This paper presents a new velocity dispersion
measure of a compact early-type galaxy (ETG) at $z\sim$1.4, and it discusses the up to now available measures.
The new $\sigma_{v}$ measure is based on near-IR LBT-LUCIFER observations. Observations and
data reduction are described in section 2. Section 3 presents details on the kinematic measures,
which are finally discussed in section 4. We assume $H_{0}=70$ km s$^{-1}$ Mpc$^{-1}$, $\Omega_{m}$=0.3
and $\Omega_{\Lambda}$=0.7. Magnitudes are in Vega system unless otherwise specified.

\section[]{Observations and data reduction}
The target is S2F1-142, a bright (K=17.6) ETG that has already been studied in details on the basis of 
its available photometric (MUNICS, \citealt{d1}) and spectroscopic (\citealt{l5}) data.
Photometric available data included the V, R, I, J and K bands, while the low resolution spectroscopic data covered
the range between 0.95 and 2.3 $\mu$m.
Furthermore, during HST CYCLE 14  we 
obtained deep HST-NICMOS high-resolution (0.075 arcsec/pix)
imaging in the F160W band ($\lambda\sim1.6$ $\mu$m) of this galaxy \citep{l7}.
The NICMOS high resolution near-IR data allowed us to measure its
effective radius $R_{e}$ and light profile in the rest-frame R-band, revealing its compact nature (\citealt{l7}, \citealt{tru}).
%Indeed, its effective radius of $R_{e\ 142}$=3.1$\pm$0.2 kpc is about 2.5 times smaller
%than the average $R_{e}$ of galaxies with similar stellar mass content, i.e.
%$\mathcal{M}_{stars}=4.1 \pm 2.0 \times 10^{11}$ M$_{\odot}$ (Chabrier IMF, \citealt{c3}).
Table 1 summarizes the basic parameters and physical properties of  S2F1-142, while
Figure 1 shows its HST F160W image  and its light profile.

\begin{table*}
\centerline{
\begin{tabular}{lccccccc}
\hline
\hline
  Object  & F160W$_{fit}^{(1)}$& M$_{R}$ & r$_e$ &  $R_{e}$ &$\langle\mu\rangle_e^{F160W}$ & $\langle\mu\rangle_e^R$
   & K \\
           &[mag]      &[mag]   & [arcsec] &  [kpc] & [mag/arcsec$^2$]& 
          [mag/arcsec$^2$] & mag \\
  \hline
S2F1-142 &18.65$\pm$0.03  & -24.00 &  0.36$\pm$0.02   & 3.1$\pm$0.2  &   20.5$\pm$0.2   &  18.4$\pm$0.2  &17.8$\pm$ 0.1   \\
\hline
\hline
\end{tabular}
}
\caption{Basic properties of S2F1-142.
$(1)$: total integrated magnitude as extrapolated on the basis of the light profile fit.} 
\end{table*}

\begin{figure}
\includegraphics[width=4cm]{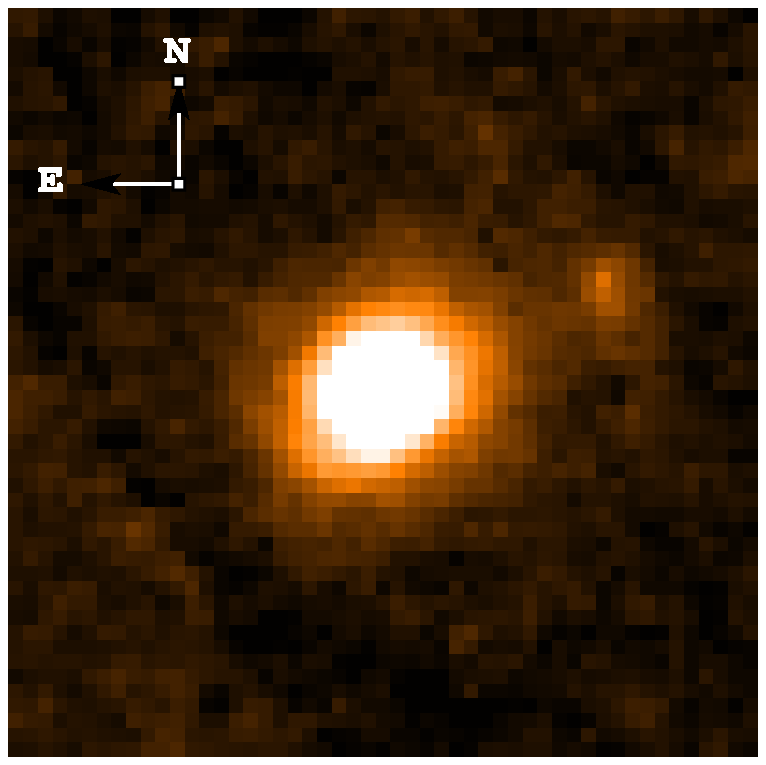}
\includegraphics[width=4.5cm]{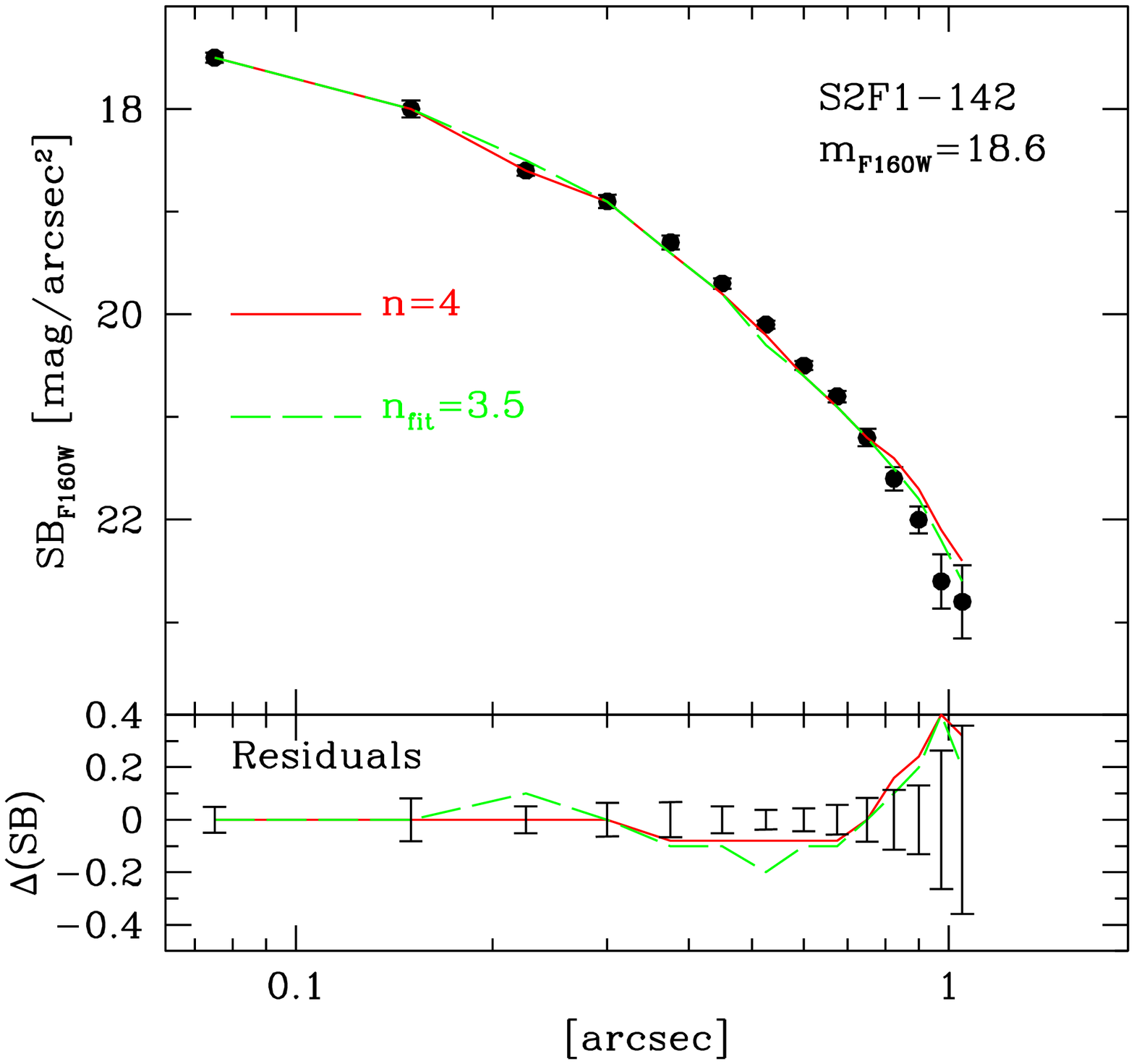}
\caption{{\it Left panel:} F160 image of S2F1-142, 3.7$^{''}$x3.7$^{''}$, total exposure time is 1h20m. {\it
Right panel:} observed (black points) and fitted (red and green lines) light profile of S2F1-142 in 
the F160 filter and residuals between observed and fitted 
profiles \citep{l7}. Green and red lines report the results obtained assuming the Sersic and De Vaucouleurs profile, respectively.  }
\end{figure}

The spectroscopic observations of the target have been carried out with LUCIFER on 
the Large Binocular Telescope (LBT) at Mt. Graham (Arizona, USA)
in long slit spectroscopic mode, during the semesters 2010A and 2010B.

\noindent
Observations during 2010A have been performed for 3 hours under good sky 
conditions (15th January and 18th February) and for other 
45 minutes under poor transparency and bad seeing conditions (15th January).
During data reduction we decided to discard the 45 minutes of observations
made under bad sky conditions, since we verified that 
they do not contribute in increasing the final signal to noise.

\noindent
The N3.75 camera was used for this first set of observations, 
coupled with the 1$\arcsec$  slit. The K$_{s}$ filter
and the grating 150\_K$_{s}$ have been adopted to cover the wavelength range 1.9$<\lambda<$2.3 $\mu$m
with a sampling of 1.3\AA\ per pixel. The resulting spectral resolution is $R\simeq$2000 
corresponding to FWHM=10\AA\ ($\sigma_{v}$=60 km/s).

\noindent
Observations carried out in 2010B added 9 hours of exposure on the target, all performed
under good sky conditions.
With respect to period 2010A, for this second set of
observations we adopted the N1.8 camera 
that coupled with the same slit, grating and filter as before, results in the
same spectral resolution but in a larger sampling (2.6\AA) and a slightly larger
wavelength coverage.

\noindent
All the observations have been performed by fixing the PA at -68.8 degree
(North to East) in order
to align the slit along the direction connecting our target S2F1-142 with a reference
bright star (SDSS J030633.97-000242.8, K=15.2). 
This observational configuration has been chosen to help
the slit positioning on our faint target. Furthermore, the bright reference star 
has been used during data reduction to correct for the final resulting sensitivity 
function (see below).

\noindent
Scientific observations have been split into many couples of short 
($\sim$ 5 minutes) exposures with the target located in two
positions (A and B) along the slit (spaced $\simeq 10$ arcsec).

\noindent
In each of the observing session we collected also some flat field images obtained with
internal quartz lamp during daytime and used to build the final frame adopted to flatfield
the scientific frames.

Images have been reduced by means of IRAF\footnote{IRAF (Image Reduction and Analysis Facility) is distributed
by the National Optical Astronomy Observatories,
which are operated by the Association of Universities for
Research in Astronomy, Inc., under cooperative agreement
with the National Science Foundation.} and following standard longslit reduction procedures.
Wavelength calibration has been performed
by means of the identification of the emission sky lines in each single frame. 
The row by row identification of the sky lines in each frame
has also supplied the correction for their curvature along the spatial
direction.  Wavelength calibration on each single frame
results correct within 0.30\AA\ and 0.15\AA\ (rms with respect to the values of the identified lines)
for the frames observed in 2010A and 2010B, respectively.
A first step of background subtraction on each single frame has been performed
by subtracting its associated frame of the dithering observing sequence A-B. The IRAF
tool {\it background} has been then used to refine the final sky subtraction of each frame,
before aligning and coadding all of them. The final bi-dimensional spectral
frame obtained in 9 hours of observations (i.e., 2010B run) is shown in Figure 2 (bottom). 
No telluric correction has been further applied, since after the previous background correction step
no evident telluric absorption was appreciable.

Before extracting the one dimensional spectrum of the target, the final co-added images have been 
corrected for the response sensitivity functions. The latter have been derived by
means of the bright reference star SDSS J030633.97-000242.8 (within the slit of all the 
scientific observations) compared with a standard spectrum of an M2 star taken from the 
Pickles Atlas \citep{p8}. The SDSS J030633.97-000242.8 star has not been officially 
classified yet, but its colours in the optical and near-IR identify it as an M star.
We choose to compare it with the M2 spectral template from the Pickles Atlas after verifying that
results obtained by means of other M stars had negligible difference (well within the flux uncertainty
due to the statistical noise) with the adopted one. 
Once reported to the same larger dispersion of 2.6\AA\ 
per pixel, the one dimensional spectra extracted from the two observing run 
have been summed to a single final spectrum. The one dimensional spectra
have been extracted just adding the rows containing signal higher than 3 $\sigma$ of noise, and
following the spatial distortion along the slit.   Figure 3 shows the final resulting spectrum 
(gray and black lines) together with the residual associated noise (gray line in the lower part of the figure): the corresponding
total S/N per pixel is $\approx$ 12 in all the regions not affected by strong emission sky lines 
corresponding to about 7 \AA\ in the restframe.

\begin{figure}
\includegraphics[width=9cm]{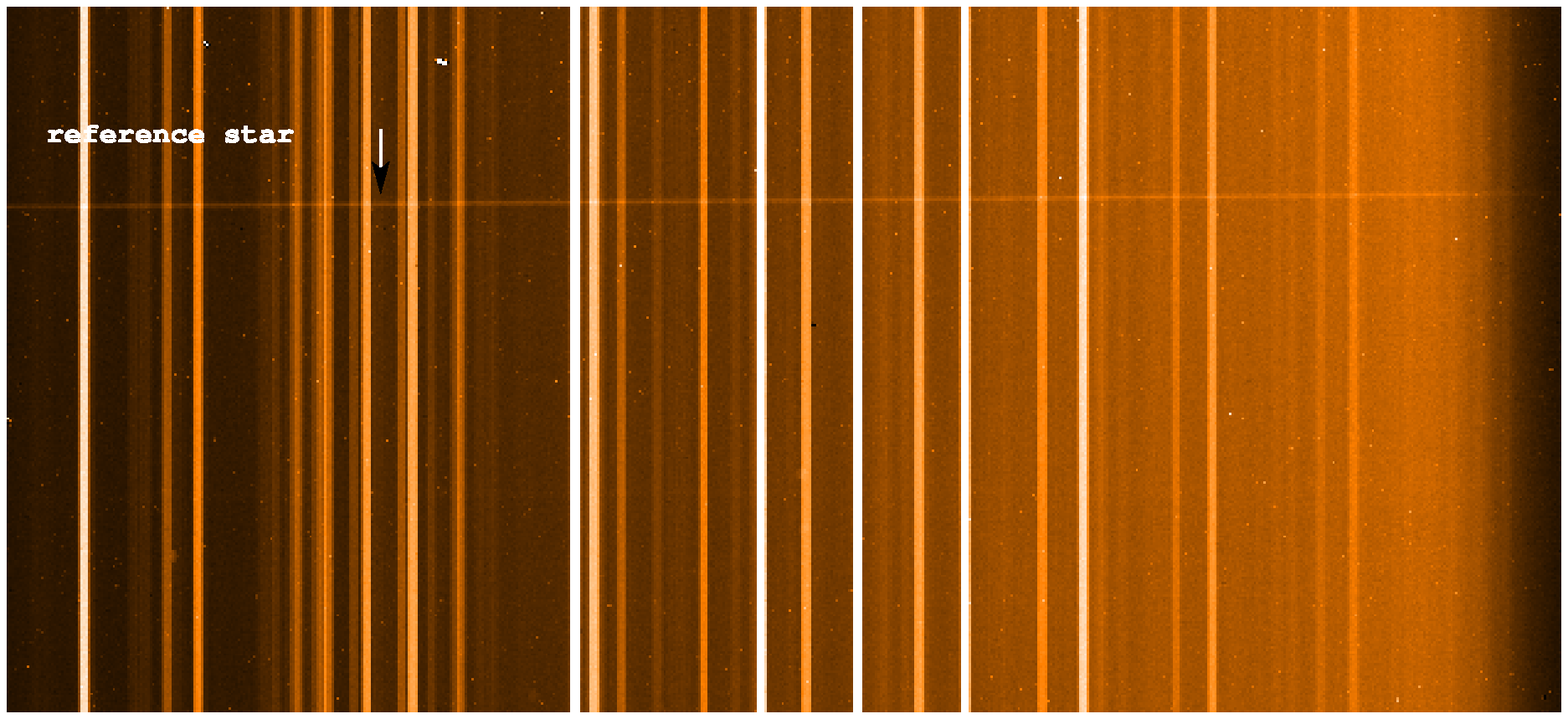}
\includegraphics[width=9cm]{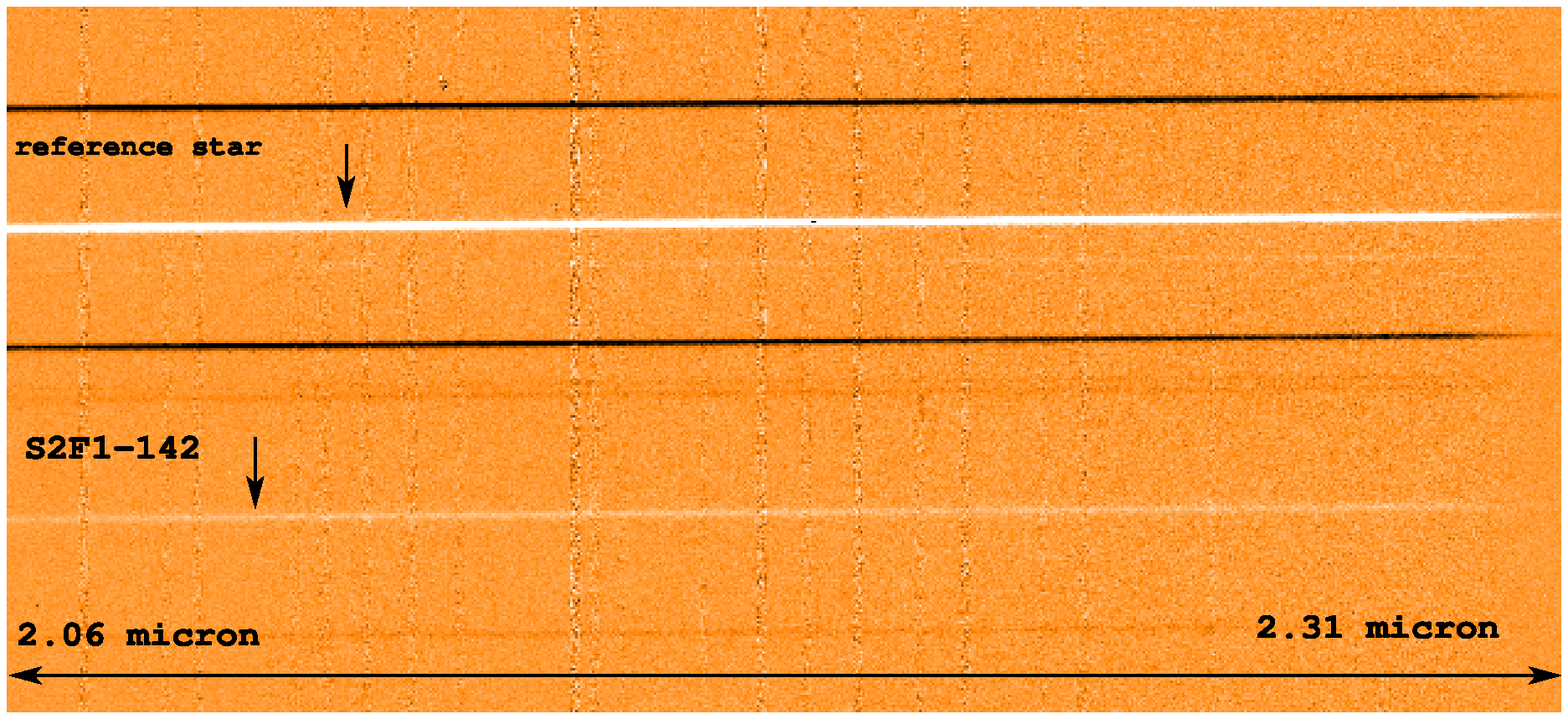}
\caption{{\it Top}: LBT-LUCIFER frame (5 minutes of exposure time). {\it Bottom}: final
reduced bi-dimensional spectrum (9 hours of exposure time).
}
\end{figure}

\begin{figure}
\includegraphics[width=8.8cm]{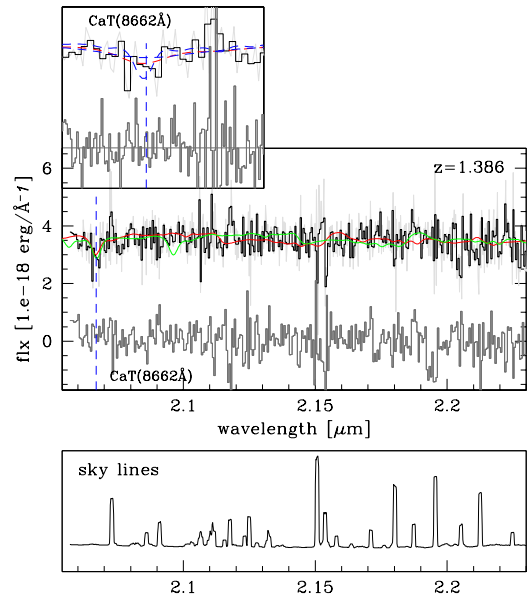}
\caption{Extracted spectrum of S2F1-142, corresponding to 12 hours of observations (light-gray line). 
The black (thick line)
histogram shows the same spectrum binned over 4 pixels ($\simeq 10$\AA). The red line
represents the bestfit template. Residual noise (i.e., the residual signal left after the sky
subtraction procedure in a background region)
 is also reported (light gray line in the bottom part of the figure). The green line shows for comparison
the same template as the red line but at $z=1.42$ as in the case the detected line was the CaT 8542\AA\ line.
In the up left corner, a zoom view of the identified CaT(8662\AA) absorption line is shown together with its best fit
modeling ($\sigma_{v}$=340 km/sec, red solid line) and other two templates corresponding to 
$\sigma_{v}$=280 and $\sigma_{v}$=460 km/sec. In the bottom box, the observed sky emission is reported for comparison.}
\end{figure}

\section{Redshift measure and velocity dispersion estimate}

As can be seen from Figure 3, only one absorption line, centered at $\lambda=$2.067$\pm 0.001 \mu$m, 
can be identified in the observed spectrum. 
The error in the line centering has been conservatively evaluated assuming 3$\sigma$ of the expected instrumental
line width.
Its position is far enough from any sky emission line (\citealt{rou})
to exclude that is a false absorption feature due to problems with the sky subtraction.
Figure 3 reports also the residual sky noise (i.e., the residual signal after the total sky subtraction step
measured in a background region) that does not show any similar absorption with
a position coincident with the identified spectral line at $\lambda=$2.067 $\mu$. We further check for possible 
telluric absorption residuals located at the same wavelength, and we could exclude
such a circumstance. 
The previous redshift estimate 
based on the very low resolution TNG-AMICI  (i.e., R=35, \citealt{l5}) data is $z=1.43 \pm 0.05$.  The large uncertainty affecting
our previous measure was due to the fact that redshift
was estimated by defining the 4000 \AA\ break position, being any other absorption feature
not appreciable at the given very low resolution. Since the 4000 \AA\ break is not a sharp absorption
feature, we considered as reliable all the redshift determination derived by a SED fitting within 1 $\sigma$ 
from the observed and available photometric and spectroscopic SED of the galaxy, that resulted to be 0.05 from the average value. 
We therefore look for the identification of the unique observed absorption line in the LUCIFER spectrum
moving $z$ within the range 1.38$<z<$1.48. Considering that the starting observed wavelength of the LUCIFER spectrum
of S2F1-142 is $\lambda=2.055 \mu$m, we identify this absorption line as the third line of the Ca triplet at 8662 \AA\ 
which gives $z=1.386 \pm 0.001$, consistently with the previous rough measure. Indeed, 
a redshift value larger than 1.40 would imply the inclusion of at least one of the other two 
CaT lines (i.e., at $\lambda$ 8542 \AA) while for $z>1.43$ all the three CaT lines (8494 \AA, 8542 \AA\ and 8662 \AA)
 would be included in the observed spectral range. As an example, in Fig. 3 (green line) a spectral template
at $z=$1.42 (i.e., for which the detected line at $\lambda=2.055 \mu$m is identified with the
CaT8542\AA\ line) is reported from which it clearly appears that the not detected 8662 \AA\ line is expected.

Figure 3 shows the good continuum matching between the 
 whole observed spectrum (thick black line) and a bestfit synthetic template (red line).
Figure 4 shows the full available SED of S2F1-142 with superimposed the best fit template
at $z=1.386$. In particular, the photometric magnitudes (green filled points) are from MUNICS (\citealt{d1}),
while the low resolution part of the SED between 0.9$\mu$m and 2.1$\mu$m is the TNG-NICS low resolution (R=35, \citealt{l5}) 
spectrum.

\begin{figure}
\includegraphics[width=8.8cm]{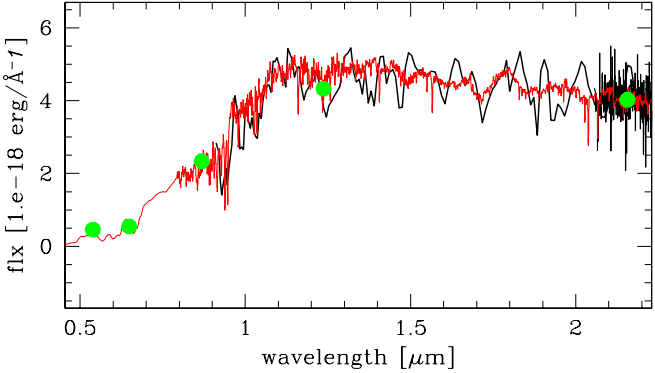}
\caption{The full available SED of S2F1-142 is shown with superimposed the best fit template
at $z=1.386$. Green filled points are the photometric magnitudes from MUNICS \citealt{d1}; 
black lines represent both the TNG-NICS low resolution (R=35, \citealt{l5}) spectrum from
0.9$\mu$m and 2.1$\mu$m and the here presented LBT 
LUCIFER spectrum between 2.0$\mu$m and 2.3$\mu$m; red line is the best fit template of the full SED at the measured redshift $z=1.386$,
built with the spectrophotometric code of \citet{bc03}, assuming solar metallicity (age=2.2 Gyr).
}
\end{figure}

The velocity dispersion estimate is based on the only available line at $\lambda=$2.067 $\mu$m.
We selected the spectral range involved by the CaT(8662\AA) line in
the observed spectrum, between 2.057$\mu$m and 2.087$\mu$m. Then we selected the same spectral range 
within a sample of synthetic templates.
Templates have been built with the spectrophotometric code of \citet{bc03}, assuming solar metallicity 
and a fixed exponentially declining star formation history with time scale of 0.1 Gyr, selecting
5 age values: 1.0, 1.6, 2.0, 3.0, 4.0 Gyr (i.e., lower than the age of the universe at $z=1.386$ that
is 4.6 Gyr). 
Each template has been redshifted at $z=1.386$. The intrinsic resolution of the \citet{bc03} models
is 3 \AA\ (FWHM, $\sigma$=50 km/sec at $\lambda \simeq 8000$\AA) 
that becomes 7.2 \AA\ at the galaxy redshift. The instrumental resolution of 
the LBT spectrum  is 10 \AA\ (FWHM). Thus we applied a gaussian convolution
with $\sigma$=3.3 \AA\ to all the synthetic redshifted templates making them equivalent
to galaxies with velocity dispersion $\sigma_{v}$=0.0 km/s observed by means of
the same instrumental setup used to observe S2F1-142. This starting set of 5 templates has thus been
built in order to simulate exactly the intrinsic instrumental resolution of our
observations.  Finally, the 5 templates  have been convolved
with gaussian curves with different $\sigma_{v}$ values 
simulating galaxy velocity dispersion between 200 and 600 km/s 
with steps of 10 km/s, resulting in a template library containing 200 templates (5 different ages x 40 different 
values of velocity dispersion).
A simple best fitting procedure has been implemented in order to choose the template that better reproduces
the CaT(8662) line observed in the LBT spectrum. We find that all the 5
set of templates with different ages give the same results, and they converge to a best fitting
$\sigma_{v}$ value of 305 (see Figure 5). 
As a more refined measure of $\sigma_{v}$,
we have perturbed within their errors the observed fluxes in the region of the CaII line,
and we have repeated the above measurement procedure on 500 perturbed spectra. Results are summarized in the histogram 
of Figure 6, and they correspond to a final average value of $\sigma_{v}=324 \pm 58$ km/s 
and a median value equal to $\sigma_{v}=340 \pm 45$ km/s. The latter has been considered our best estimate
of $\sigma_{v}$.

In order to give a more reliable estimate of the uncertainty of our
$\sigma_{v}$ measure, we produced a set of simulations aimed at determining the probability to obtain 
the measured value $\sigma_{v}=340$ starting from  `true' different values.
We created a set of mock spectra starting from a stellar population of 2.5 Gyr (i.e.,
in agreement with the S2F1-142 age, see Table 2), redshifted at $z=1.386$
and reported to the same grid of velocity dispersions adopted in the best fitting procedure (40 templates
differing for the values of their velocity dispersion between 200 and 600 km/s).
 We then extracted 60 spectra of residual noise from 60
independent regions of the final reduced bidimensional LBT image. We added each of the noise spectra to
each of the templates in the mock catalog, resulting in a mock spectral library of 2400 spectra (40 different
templates x 60 different noise signal). We analyzed the whole mock library adopting the same procedure
adopted for the observed spectrum, thus producing a grid of `measured' $\sigma_{v_{m}}$ values versus `true' $\sigma_{v_{t}}$ known ones.
Finally, we considered the distribution of the `true' $\sigma_{v_{t}}$ values of all the mock spectra that had
the `measured' ones equal to the 340 km/sec (i.e., the S2F1-142 measured velocity dispersion).
We found that 68\% of $\sigma_{v_{t}}$ for which $\sigma_{v_{m}}=340$ km/sec
fall between 280 and 460 km/sec, i.e., $\sigma_{v}=340 ^{-60}_{+120}$ km/sec.
We than more properly assume that the measured value of the velocity dispersion of S2F1-142 $\sigma_{v}=324$ km/sec
is affected by 1$\sigma$ error of $^{-60}_{+120}$ km/sec. 
This measured value has to be referred to 1 arcsec central part of the galaxy, that
corresponds to about $1R_{e}$ circular aperture measure. In the following section, $\sigma{v}$ will
be compared with other ETGs $\sigma_{v}$ estimates from literature which in some cases are referred to
the very central portion of the galaxies, i.e., $R_{e}/8$. We therefore calculate $\sigma_{v(re/8)}=1.15\ \sigma_{v}= 391$
km/sec, following \citet{ca06}.

\begin{table*}
\centerline{
\begin{tabular}{lcccccccc}
\hline
\hline
  Object  &\multicolumn{2}{c}{Salpeter} &\multicolumn{2}{c}{$s=2.0$} &\multicolumn{2}{c}{Chabrier} &\multicolumn{2}{c}{$s=1.5$} \\
          &  Age  & $\mathcal{M}_*$ &  Age  & $\mathcal{M}_*$  &  Age  & $\mathcal{M}_*$  &  Age  & $\mathcal{M}_*$  \\
          & [Gyr] & [10$^{11}$ M$_{\odot}$] & [Gyr] & [10$^{11}$ M$_{\odot}$] & [Gyr] & [10$^{11}$ M$_{\odot}$] & [Gyr] & [10$^{11}$ M$_{\odot}$] \\
  \hline
S2F1-142 & 2.2 & 6.1 & 2.6 & 3.8 & 2.2 & 3.5 & 1.4 & 1.7 \\
\hline
\hline
\end{tabular}
}
\caption{Revised ages and masses derived by means of the SED fitting of the galaxy for different IMFs (see text). 
}
\end{table*}

\begin{figure}
\includegraphics[width=8cm]{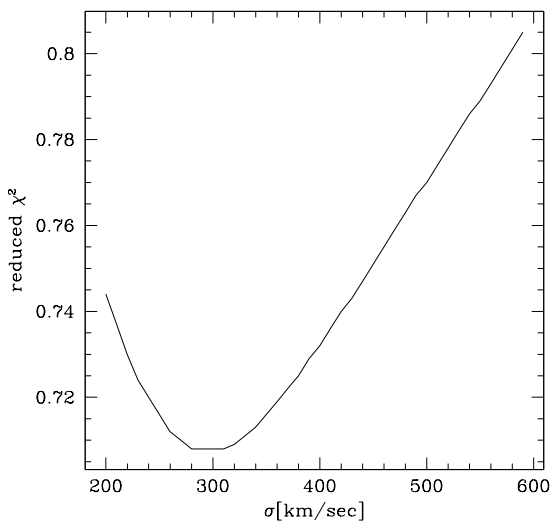}
\caption{Reduced $\chi^{2}$ distribution obtained in the CaT(8662\AA) line best fitting procedure 
assuming the set of 2 Gyrs old templates (see section 3 for more details).
}
\end{figure}

\begin{figure}
\includegraphics[width=8cm]{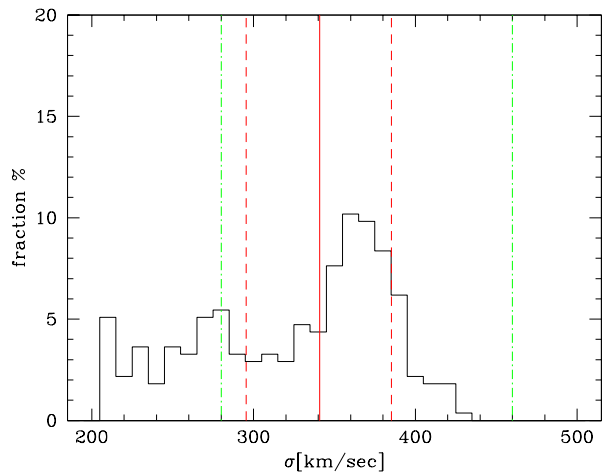}
\caption{Histogram distribution of the measured values of $\sigma_{v}$ on 500 simulated spectra
obtained by perturbing within the errors the observed fluxes in the region of the CaII line.
The red solid line represents the median of this distribution ($\sigma_{v}=340$ km/sec and the red dashed lines mark
the range including 68\% of the obtained values (i.e., $\pm 45$ km/sec). The green dot-dashed lines mark
the finally accepted range of uncertainty between $\sigma_{v}=280$ km/sec and $\sigma_{v}=460$ km/sec.
}
\end{figure}

\begin{figure*}
\includegraphics[width=8.2cm]{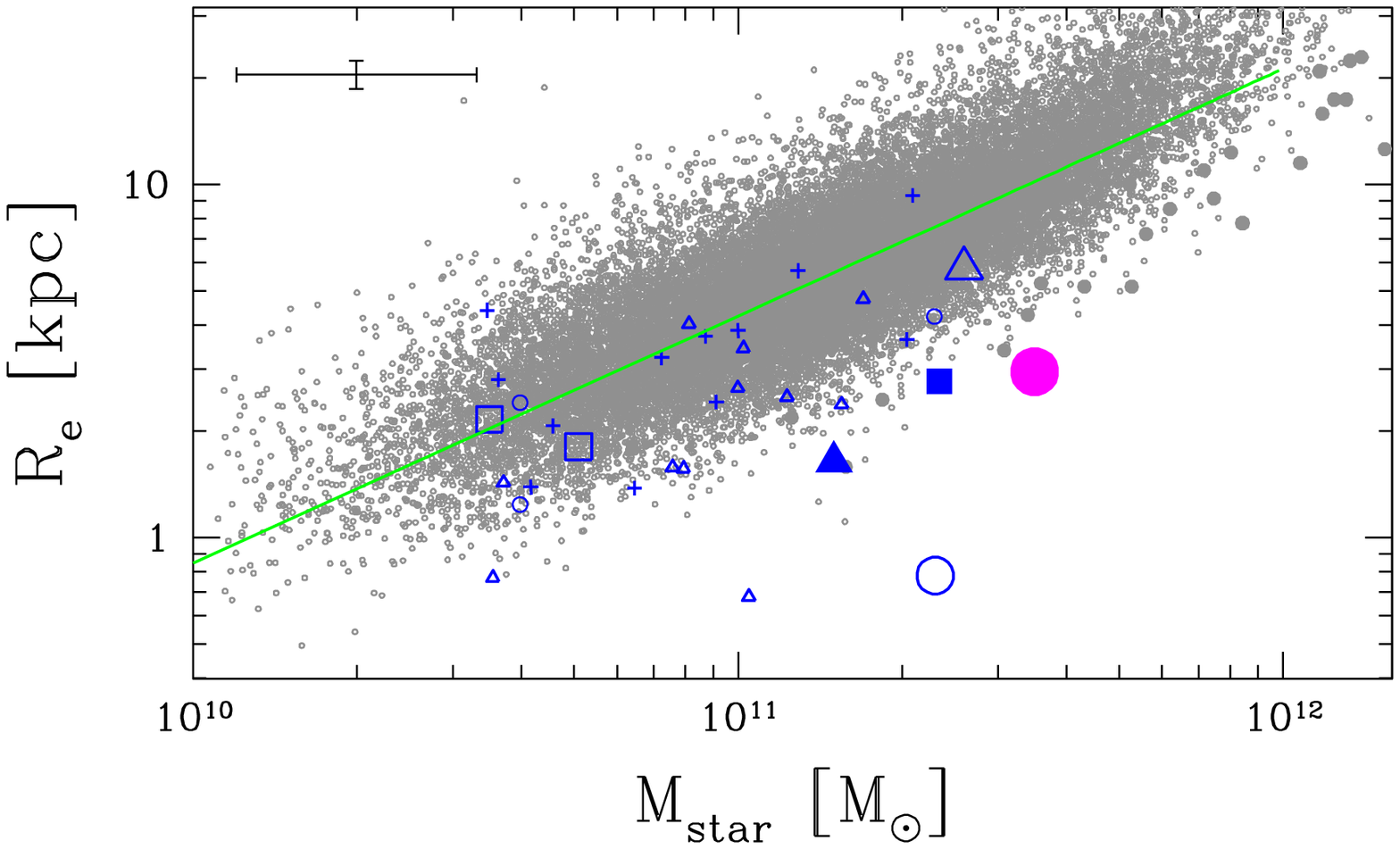}
\includegraphics[width=8.2cm]{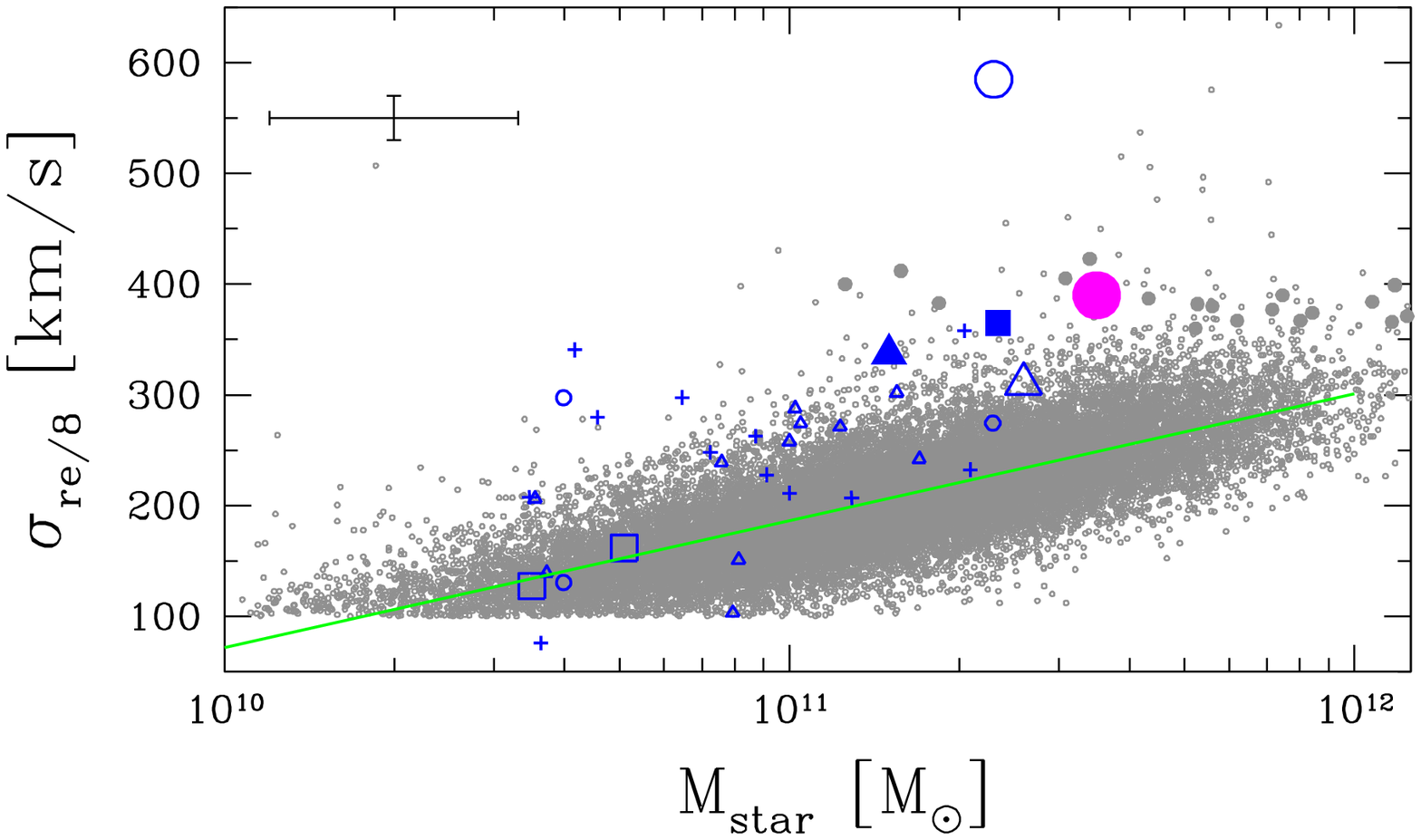}
\caption{{\it left panel:} The local size-mass distribution (gray circles) is reported
together with the line representing its average value (green solid line)
(\citealt{me13}, \citealt{ber13}).
S2F1-142 is reported as a big magenta filled circle. The other blue points represent
all the up to now available measures for galaxies at $z\ga 1$ from \citet{ono} (big open triangle), \citet{ca09} (big open squares),
\citet{vdk} (big open circle), \citet{toft} (big filled square), \citet{vds} (big filled triangle), \citet{vdw} (crosses),
\citet{dsa} (small open circles), \citet{n10} (small open triangles). Small gray squares are local measurements from
\citet{ber} of brightest cluster members with $\sigma_{gal} > 350$ km/s;
{\it right panel} the local $\sigma_{v}$-mass distribution (gray small circles) is reported
together with the line representing its average value (solid line) (\citealt{me13}, \citealt{ber13}).
Symbols are as in the left panel.
In the upper left corners of both the two panels, representative errors of the reported quantities
are shown.
}
\end{figure*}

\noindent
From the stellar velocity dispersion we calculate the virial mass as $M_{vir} = c r_{e} \sigma^{2}/G$, 
where we assume c=5 from the local calibration by \citet{ca06}. The effective radius has been measured
by \citet{l7} in the NICMOS F160W band (corresponding to the restframe R band), being $r_{e}=0.36\pm0.02$ 
arcsec that at the redshift of $z=1.386$ corresponds to $R_{e}=3.1\pm0.2$ kpc.
Thus, we estimate a virial mass of 3.6$\times10^{11}$ M$_\odot$ ($\pm2.5\times10^{11}$M$_\odot$).  
For comparison, in Table 2 the stellar mass estimates of S2F1-142 obtained by means of its SED fitting 
assuming different IMFs are reported. 
With respect to the previous value reported for the Chabrier IMF in Table 2 of \citet{s9},
estimates are slightly changed due to a better fixing of the redshift obtained in this work and to the adding of
the now available photometric near-IR measure in the WISE RSR-W1 filter (i.e.,  $\lambda_{eff} \simeq 3.4 \mu m$).
The SED bestfitting has been made with a wide set of templates derived from the \citet{bc03} models at
solar metallicity, assuming exponentially declining star formation histories with $\tau$ between 0.1 and 0.6 Gyr
and dust extinction $A_{V}$ between 0.0 and 0.6 mag. Besides the most commonly adopted IMFs by
\citet{sal} and \citet{c3}, we obtained the bestfitting parameters considering also other initial mass functions
which follow the power law scaling $dn/dm \sim m^{-s}$, with $s$ between 1.5 and 3.5. All the bestfitting solutions require
a dust extinction value $A_{V}\simeq0.5$, and the resulting {\it age} and $\mathcal{M}_*$ are reported in Table 2,
moving from dwarf richer mass functions to less steep ones from left to right. 

%The above calculated virial mass results compatible (within 1 $\sigma$ of uncertainty) only with
%$\mathcal{M}_*$ derived assuming IMFs less steep than $s=2.0$.

\section{Discussion and conclusions}
We obtained a medium resolution (R=2000) LBT-LUCIFER spectrum of the compact early-type galaxy
 S2F1-142 at $z\simeq1.4$.
In the observed wavelength range, between 2.0 $\mu$m and 2.3 $\mu$m, we detected a single absorption 
line at 2.0668 $\mu m$ that has been identified as the third
CaT line at $8662$ \AA. By means of this line we measured $z=1.386 \pm 0.001$ and we estimated 
$\sigma_{v}=340 ^{-60}-{+120}$ km/s. The corresponding virial mass has been calculated as 3.9$\times10^{11}$ M$_\odot$,
compatible with the stellar mass estimates obtained assuming IMFs less dwarf reach than a Salpeter one.

\noindent 
In Figure 7 (left panel) our target galaxy S2F1-142 is located on the size-mass diagram (filled magenta circle),
together with the distribution of ETGs in the local Universe (gray circles,  \citealt{me13}, \citealt{ber13})
and other up to now available measures for galaxies at $z\ga 1$ (blue symbols). All the mass estimates have been
determined on the basis of a \citet{c3} IMF.
S2F1-142 is a compact galaxy with an effective radius more than 3 times smaller than expected from the average
local value of galaxies with the same stellar mass content, i.e. $<R_{e}>(3.5\times 10^{11}$ M$_\odot$) = 10.2 kpc.  
In the right panel, the local distribution of the velocity dispersion values of ETGs in the local universe is reported
together with the average values (solid line) and with the same sample of high redshift galaxies reported in the
left panel. It is noticeable the large scatter of both the relations. 
First of all, we try to look for a local 
counterpart of S2F1-142, that is a local galaxy with the same mass, effective radius and velocity dispersion. 
In the sample of \citet{ber13}, we find about 20 galaxies with stellar mass and $R_{e}$ compatible with 
those of S2F1-142 (i.e., $\mathcal{M}_{stars}$ within 2 and 6 $\times10^{11}$ M$_\odot$ and $R_{e}$
between 2.5 and 3.5 kpc).
Their velocity dispersion values are within 170 and 380 km/s, thus compatible
within 1 $\sigma$ error with $\sigma_{v(re/8)}$. This means that S2F1-142 can be considered an extreme object 
but still similar to other extreme objects populating the local universe.
Finally, we can compare S2F1-142 with other high redshift galaxies. 
With respect to the galaxy presented in \citet{ono}, S2F1-142 has a a similar stellar mass content
(i.e. $\mathcal{M}_{* ONO} $ is 2.6x10$^{11}$M$_\odot$ that is  0.6 times $\mathcal{M}_{*}$ of S2F1-142), while its effective radius
is 1.9 times smaller than $R_{e\ ONO}$, where $R_{e\ ONO}=5.79$ kpc.
In case both the two galaxies were virialized, we than expect that S2F1-142 has a velocity dispersion
$\sigma_{v}=1.4 \ \sigma_{v\ ONO}$, that is a value of its velocity dispersion of 378 km/sec, compatible with the
measured $\sigma_{v}$=340$^{-60}_{+120}$ km/sec.  We obtain a similar result even if we consider the 2 galaxies from 
\citet{ca09} which have a larger difference in mass with respect to the previous case (i.e. 
$M_{*\ CAP}=3.5-5.1 x 10^11 $M$_\odot$, that are 0.08-0.12 times the stellar mass of S2F1-142) 
but similar effective radii (i.e. $R_{e\ CAP}=2.16-1.81$ that are 0.7-0.6 times
the effective radius of S2F1-142).
In this case, the expected value of $\sigma_{v}$ for S2F1-142 is 2.9-2.2 the values of the two galaxies
of \citet{ca09}, that is $\simeq$ 330 km/sec, again
compatible with the measured valued  $\sigma_{v}$=340$^{-60}_{+120}$ km/sec.

\vskip 0.5 truecm
As a final test, we check if those galaxies which have a quite lower value of $R_{e}$ than expected
from their stellar mass content and from the {\it average} local size-mass relation (like S2F1-142) have
a velocity dispersion correspondingly higher with respect to the {\it average} sigma-mass relation,
where the correspondence is evaluated in the hypothesis of virialization.

In fact, in a virialized system, it is expected that $\sigma_{v}$, $R_{e}$ and M$_{star}$ are related in the following way:

$$M_{vir}=f M_{star}=c \frac{R_{e} \sigma^{2}}{G} $$

\noindent
where $c$ is the virial coefficient (e.g., \citealt{ca06}) and $f$ is the ratio between  the total virial mass
and the stellar mass.
Assuming $\frac{f}{c}$  constant, we expect
that the velocity dispersion of two galaxies with different stellar mass content M$_{star 1}$ and
M$_{star 2}$, and different effective radii $R_{e 1}$ and $R_{e 2}$ relate themselves as:
$$\frac{\sigma_{1}}{\sigma_{2}} =\sqrt{ \frac{M_{star 1}}{R_{e 1}}  \frac{R_{e 2}}{M_{star 2}}}$$

\noindent
Thus, a galaxy like S2F1-142, whose $R_{e}$ is more than three times the average local value, is expected to have
a velocity dispersion almost twice the mean local value of ETGs with its stellar mass.
From the relation in Figure 7 (right panel), the local value of $\sigma_{v}$
for galaxies with stellar mass equal to 3.5$\times10^{11}$ M$_\odot$ is 250 km/s (within a circular aperture of
$R_{e}$/8), that is only 1.5 times smaller
than $\sigma_{v(re/8)}$ of S2F1-142-like galaxies  (1.7 considering 1$\sigma$ error).
To further check this evidence,  
in Figure 8, we report $\sigma_{v}/\sigma_{v\ z=0}$  versus $R_{e}/R_{e\ z=0}$, where the values of $\sigma_{v\ z=0}$
and $R_{e\ z=0}$ are the average local values (green lines in Figure 7, left and right panel, respectively) 
for galaxies with the same stellar mass content.
Symbols are the same as in previous Figure 7. It is immediately noticeable
that the distribution
of these quantities does not appear to depend on the redshift. 
The best fitting of the whole set of data results to be

$$\frac{\sigma_{1}}{\sigma_{2}} =[ \frac{M_{star 1}}{R_{e 1}}  \frac{R_{e 2}}{M_{star 2}}]^{\alpha}$$

\noindent
where $\alpha$ is 0.20, and not 0.50 as expected on the basis of the previous hypothesis.
The easiest way to explain this finding is to release the hypothesis that $\frac{c}{f}$ is constant for all the ETGs,
assuming on the contrary that it depends on their compactness. If this would be the case, the result $\alpha=0.2$ 
translates into $\frac{f}{c} \sim [\frac{M_{star}}{R_{e}}]^{-0.60}$.

\noindent
As pure speculation, we make the hypothesis that
the virial coefficient $c$ is not systematically
dependent on the compactness of galaxies. The above finding 
can then be read as if at fixed mass the dark matter fraction $f$ is higher in more diffuse galaxies 
while it is smaller in the most compact ones. Compact galaxies, both at high redshift and in the local
universe, could have a lower dark matter content with respect to their more diffuse counterparts
with the same stellar mass. This evidence would be consistent with the dry mergers scenario,
where the more relaxed galaxies are the product of stars stripping from low mass satellites by more compact objects,
which assemble the stellar particles of the satellite at large radii in halo-dominated
regions of the massive host.
 This process strongly increases the size of the bulge into regions with higher dark matter fractions,
 leaving the inner host structure almost unchanged
\citet{hil12}.
At the same time, this process does not seem an evolutionary effect, since the consequent relation
between dark matter content and compactness is observed at any redshift.

\vskip 0.3 truecm
On the other hand, making the opposite hypothesis than before, that is the dark matter
fraction in ETGs is constant and not dependent on their compactness, the previous finding would translate into 
${c} \sim [\frac{M_{star}}{R_{e}}]^{+0.60}$. The virial coefficient $c$, at least for an isothermal sphere model,
increases at increasing values on the profile Sersic fitting parameter $n$ (\citealt{bert}), 
so that we can see $c$ almost proportional to $n$. This means that under the hypothesis of
fixed dark matter fraction, the above finding translates into an expected dependence of the luminosity profile
steepness on the compactness of the galaxies. This, differently from the previous case, would be in contrast
with the inside-out dry mergers models, that foresees a larger value of $n$ for the less compact evolved ETGs (\citealt{pat13}).

\vskip 0.3 truecm

Finally,
another simpler explanation of our finding of a high value of $\alpha$ in  $\frac{\sigma_{1}}{\sigma_{2}} =[ \frac{M_{star 1}}{R_{e 1}}  
\frac{R_{e 2}}{M_{star 2}}]^{\alpha}$, without invoking varying dark matter content or non homology,
could be that the stellar mass content of the apparently more compact galaxies is  overestimated, and their
real compactness is thus less extreme than assumed. A possible reason of a systematic overestimate of stellar masses 
of the more compact galaxies could be related to the IMF assumed to transform their luminosity into
a mass estimate. 
%Indeed,  bottom light IMF, 
%such as the one expressed by a power law 
%with the exponent $s=1.5$ (see Table 2), would reduce the stellar mass content of S2F1-142 to the value of $\simeq 2\times10^{11}$ M$_\odot$
%for which we have above noticed that its expected velocity dispersion value would be in agreement
%with the measured one. 
If this would be the case, we can expect that many of the up to now claimed compact ETGs
are actually more relaxed galaxies for which the stellar mass estimate has been overestimated by
assuming a too steep IMF (i.e., bottom heavy IMF). It is worth to note that evidences of a non universal IMF have 
been recently found out by many authors (e.g., \citealt{ca12}, \citealt{con13}, \citealt{sl13}) even if a common
view of its possible variation with density is far to be reached. In particular, while \citet{con13}
find evidences that favor a steeper IMF in more massive ETGs,
\citet{sl13} demonstrated the necessity of a bottom light IMF in a particularly massive elliptical galaxy at $z=2.14$.
Under this last hypothesis, our finding on the connection  between the average size-mass relation and the average sigma-mass relation
suggests the possibility that the stellar mass of those galaxies which appear much denser than the average relation  is lower
than deduced by assuming a universal IMF,
thus better supporting the \citealt{sl13} finding.
It is anyhow worth to note that stellar mass estimates are often affected by large
uncertainties not only related to the assumed IMF, but also to the uncertain age and metallicity of their
stellar populations, and to the inaccuracy of the stellar populations models themselves.

\vskip 0.5truecm
Summarizing, S2F1-142 is a compact galaxy that is fully consistent with the already available measures
of $R_{e}$ and $\sigma_{v}$
both in the local and in the distant universe.
%At the same time, its velocity dispersion is measured lower than expected from its apparent
%compactness and on the basis of the local distributions of $\sigma_{v}$ and $R_{e}$.
%We propose that S2F1-142 has a lower dark matter content with respect to the more relaxed
%ETGs in the local Universe, and/or, 
%its larger compactness is an apparent effect due to an intrinsic IMF less dwarf rich
%than the one typical of the local distribution.
Furthermore, both in the local and in the distant universe, we note evidence that the size-stellar mass
relation and the $\sigma$-stellar mass one cannot be simply related to each other by assuming the virial
theorem with an average universal dark matter fraction and/or a strict homology of all the ETGs. The latter
two quantities should depend on the galaxies compactness. Alternatively, 
many of the up to now claimed compact ETGs
are actually more relaxed galaxies for which the stellar mass has been overestimated, for example because
of the assumption of a too steep IMF (i.e., bottom heavy IMF), or alternatively, because
of some stellar parameters systematic. 

\begin{figure}
\includegraphics[width=8.2cm]{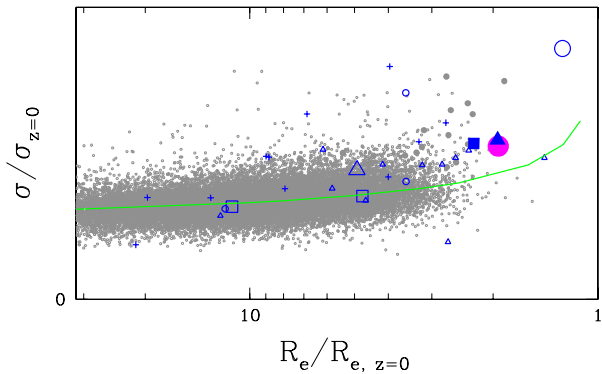}
\caption{For the same samples of previous Figure 7 (reported with the same symbols), the distribution of $\sigma/\sigma_{z=0}$ values
is reported versus $R_{e}/R_{e\ z=0}$ values, where $\sigma_{z=0}$ and $R_{e\ z=0}$ are the average local values
as derived from the relations in Figure 7 (green lines).
}
\end{figure}

\section*{Acknowledgments}
This work is based on observations made at the Large Binocular Telescope (LBT) at Mt. Graham (Arizona, USA).
The LBT is an international
collaboration among institutions in the United States, Italy and
Germany.
LBT Corporation partners are the University of Arizona on
behalf of the Arizona university system; Istituto Nazionale di
Astrofisica (INAF), Italy; LBT Beteiligungsgesellschaft, Germany,
representing the Max-Planck Society, the Astrophysical Institute
Potsdam, and Heidelberg University; the Ohio State University, and the
Research Corporation, on behalf of the University of Notre Dame,
University of Minnesota and University of Virginia.
We thank F. Mannucci and the LBT team for their help in both preparing and conducting the observations.
We are grateful to M. Bernardi and her team for providing unpublished data of their local
samples, and to S. Charlot for providing models with different Initial Mass Functions.
We also thank the anonymous referee for her/his helpful comments which greatly improved 
the clarity of this manuscript.
This work has received financial support from Prin-INAF
1.05.09.01.05

\label{lastpage}

\end{document}